\documentclass[superscriptaddress,
  aps,
  pra,
  amsmath,amssymb,
  reprint,
]{revtex4-1}
\usepackage{graphicx,color}
\usepackage{amsmath}
\usepackage{natbib}
\usepackage{epsfig}
\usepackage{graphicx}

\begin{document}

\title{\color{blue} Dispersion relations of Yukawa fluids at weak and moderate coupling}

\author{Sergey Khrapak}
\email{Sergey.Khrapak@dlr.de}
\affiliation{Institut f\"ur Materialphysik im Weltraum, Deutsches Zentrum f\"ur Luft- und Raumfahrt (DLR), 82234 We{\ss}ling, Germany}
\affiliation{Joint Institute for High Temperatures, Russian Academy of Sciences, 125412 Moscow, Russia}

\author{L{\'e}na{\"ic} Cou\"edel}
\affiliation{Physics and Engineering Physics Department, University of Saskatchewan, 116 Science Place, S7N 5E2 Saskatoon, Saskatchewan, Canada}
\affiliation{CNRS, Aix-Marseille Universit{\'e}, Laboratoire PIIM, UMR 7345, 13397 Marseille cedex 20, France}

\begin{abstract}
In this paper we compare different theoretical approaches to describe the dispersion of collective modes in Yukawa fluids when the inter-particle coupling is relatively weak, so that kinetic and potential contributions to the dispersion relation compete. Thorough comparison with the results from molecular dymamics simulation allows us to conclude that in the regime investigated the best description is provided by the sum of the generalized excess bulk modulus and the Bohm-Gross kinetic term.      
\end{abstract}

\date{\today}

\maketitle

\section{Introduction}

There has been substantial evidence that the quasi-localized charge approximation (QLCA), also known as the quasi-crystalline approximation (QCA), describes rather well the long-wavelength portion of the dispersion relations of collective excitations in strongly coupled Yukawa fluids~\cite{RosenbergPRE1997,OhtaPRL2000,KalmanPRL2000,DonkoJPCM2008,
KhrapakPoP2016,KhrapakIEEE2018}, including the one-component plasma (OCP) limit~\cite{KaGo1990,GoldenPoP2000,KhrapakAIPAdv2017,KhrapakJCP2018}. The purpose of this study is to answer the question of what determines the dispersion relation of Yukawa fluids at moderate coupling.

This regime corresponds to the ``true'' fluid situation, where no small parameter is present. At weak coupling, interactions between particles provide a small correction to the conventional multicomponent plasma dispersion relation. At strong coupling, QLCA and QCA do a rather good job and kinetic corrections are numerically small and can normally be neglected. What can be an appropriate theoretical approximation in between these two limits?      

We provide answer to this question below. Since well defined transverse (shear) modes in fluids are normally supported only in the vicinity of the fluid-solid phase transition (strong coupling regime)~\cite{OhtaPRL2000,NosenkoPRL2006,GoreePRE2012,OttPRE2013,KhrapakJCP2019} we concentrate on the longitudinal mode here. The transverse mode will be mentioned only briefly, to the extent necessary for the understanding of the proposed approximations. 

Recently, the evolution of the longitudinal sound velocity of Yukawa systems from the weak- through the strong-coupling regimes has been studied in detail in Ref.~\cite{SilvestriPRE2019}. Sound velocity can be related to thermodynamic quantities and hence knowledge of an appropriate equation of state can solve the problem. Here we analyse the entire dispersion curves, not only their long-wavelength asymptotes.   

Extensive molecular dynamics (MD) simulations have been performed to obtain dispersion relations of weakly and moderately coupled Yukawa fluids. Theoretical approximations applicable to this regime are discussed and the most suitable is identified. It turns out that the generalized excess bulk modulus supplemented by the Bohm-Gross kinetic term provides particularly good theoretical description of the numerically obtained dispersion curves.     

\section{Yukawa fluids}

Historically, interest to classical systems of particles interacting via the repulsive Yukawa  (screened Coulomb or Debye-H\"uckel) potential was mainly related to modeling charges immersed in a polarizable background, e.g. electron-ion plasma and charge-stabilized colloidal dispersions~\cite{BarratJPCSSP1988,RobbinsJCP1988,IvlevBook}. More recently, Yukawa potential has been extensively used as a first approximation to model interactions between macroscopic particles in complex (dusty) plasmas~\cite{TsytovichUFN1997,FortovUFN,FortovPR,KhrapakPRL2008,FortovBook,ChaudhuriSM2011}. In a more general context, the Yukawa potential represents an important example of soft repulsive interactions operating in various soft matter systems.   

In Yukawa systems particles are interacting via the pairwise potential of the form 
\begin{equation}\label{Yukawa}
\phi(r)=(Q^2/r)\exp(-r/\lambda),
\end{equation}
where $Q$ is the particle charge and $\lambda$ is the screening length. Such a system is fully characterized by the two dimensionless parameters: the coupling parameter $\Gamma=Q^2/aT$ and the screening parameter $\kappa=a/\lambda$, where $a=(4\pi n/3)^{-1/3}$ is the Wigner-Seitz radius,  $T$ is the temperature in energy units ($k_{\rm B}=1$), and $n$ is the density. Conventionally, the system is referred to as strongly coupled (non-ideal) when $\Gamma\gg 1$, that is when the Coulomb interaction energy exceeds considerably the kinetic energy (more precisely, when $\phi(a)\gg T$, so that screening is accounted for). The opposite limit $\Gamma\ll 1$ corresponds to the weakly coupled (ideal) regime.

\begin{figure}
\includegraphics[width=7cm]{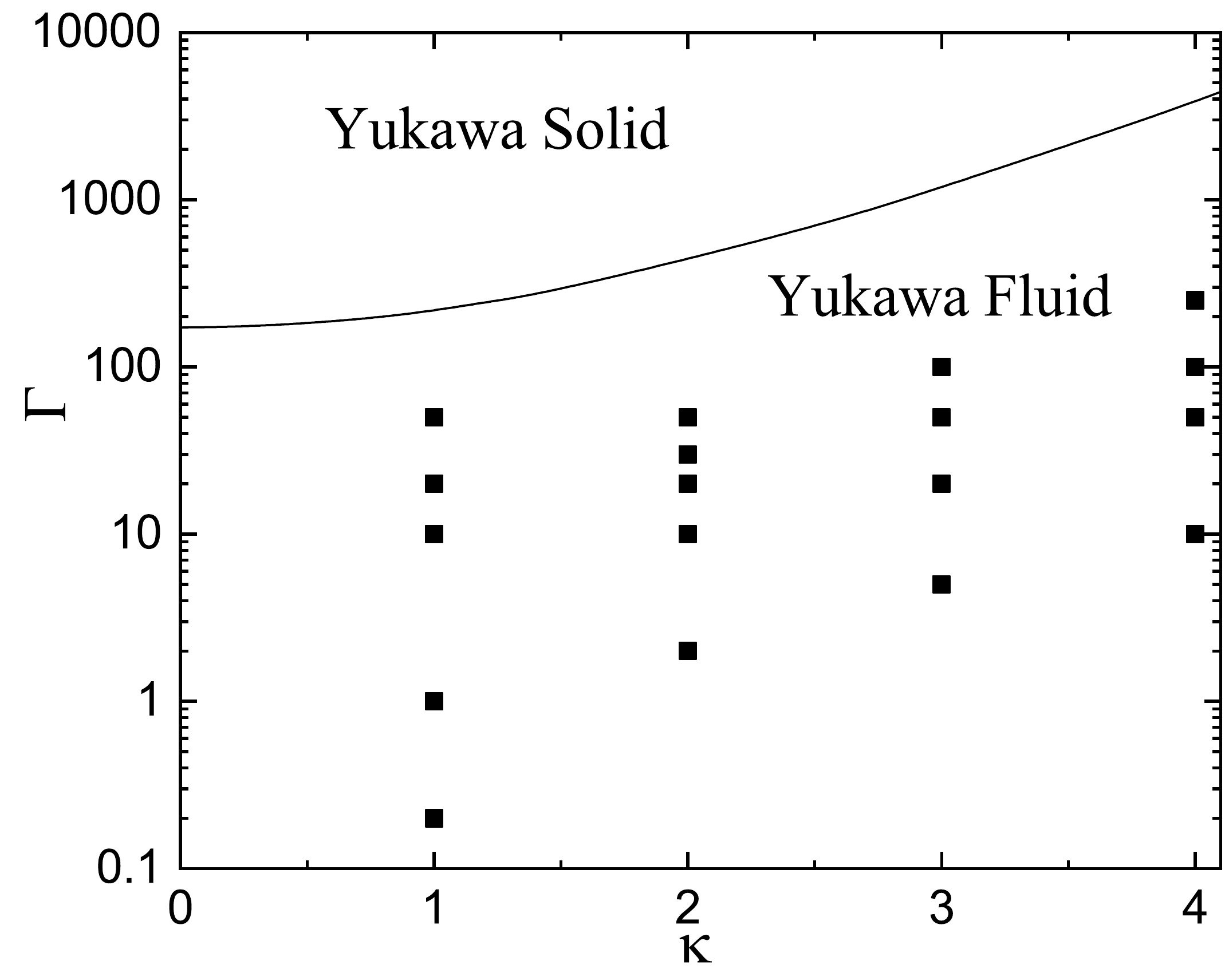}
\caption{Phase diagram of Yukawa systems in a ($\kappa$, $\Gamma$) plane. The solid curve corresponds to the fluid-solid (melting) phase transition according to the data from Ref.~\cite{HamaguchiPRE1997}. Symbols correspond to the phase state points investigated in this work. }
\label{Fig1}
\end{figure}

The phase diagram of theree-dimensional Yukawa systems in ($\kappa$, $\Gamma$) plane is shown in Fig.~\ref{Fig1}. The solid curve smoothly connect the melting points data obtained from the free energy consideration and tabulated in Ref.~\cite{HamaguchiPRE1997} (accurate analytical fits are also available~\cite{VaulinaJETP2000,VaulinaPRE2002}). We consider moderate screening regime with $1\leq \kappa \leq 4$, which is particularly relevant for complex plasma experiments in gas discharges. In the considered range of $\kappa$, a Yukawa fluid first freezes into the bcc lattice. The fcc lattice can be also stable, but either at higher $\kappa$ or at higher $\Gamma$ (only for $\kappa\gtrsim 1$)~\cite{HamaguchiPRE1997}. It should be noted that in real experiments with complex plasmas, a metastable hcp lattice can be present or even be a dominant constituent of the solid phase~\cite{NefedovNJP2003,KhrapakPRL2011,KhrapakPRE2012}. Symbols in the theoretical phase diagram depicted in Fig.~\ref{Fig1} correspond to state points investigated in this work. 

\section{Theoretical approaches and strategy}

\subsection{Fluid description}

We start with a minimalistic simple description of multi-component plasmas, similar to that used in the original derivation of the dust-acoustic wave (DAW) dispersion relation in Ref.~\cite{RaoDAW}. In this formulation electrons and ions provide equilibrium neutralizing medium and follow the Boltzmann distribution in the wave potential: 
\begin{equation}\label{Boltzmann}
n_i\simeq n_{i0}e^{-e\varphi/T_i},\quad\quad n_e\simeq n_{e0}e^{e\varphi/T_e},
\end{equation}
where $\varphi$ is the electric potential, $e$ is the elementary charge (ions are assumed singly charged), $n_{i0}$ and $n_{e0}$ are the unperturbed ion and electron densities, and $T_i$ and $T_e$ are their respective temperatures (again in energy units).  

The continuity and momentum equation for the (dust) particle component are
\begin{equation}\label{continuity}
\frac{\partial n}{\partial t}+\nabla(n{\bf v})=0,
\end{equation}
\begin{equation}\label{momentum}
\frac{\partial {\bf v}}{\partial t}+({\bf v}\cdot \nabla){\bf v}=-\frac{Q\nabla\varphi}{m}-\frac{\nabla P}{mn},
\end{equation}
where $P$ is the pressure associated with the particle component. 

The system is closed by the Poisson equation,
\begin{equation}\label{Poisson}
\Delta \varphi = -4\pi\left(en_i-en_e+Qn\right).
\end{equation}
The particle charge is assumed fixed for simplicity.

Linearization of the systems of equations (\ref{Boltzmann})-(\ref{Poisson}) constitutes a standard exercise in plasma physics. The result can be written in the following form~\cite{FortovPR}
\begin{equation}\label{DAW}
\omega^2 = \frac{\omega_{\rm p}^2 k^2\lambda^2}{1+k^2\lambda^2}+\mu k^2 v_{\rm T}^2,
\end{equation}  
where $\omega$ is the frequency, $k$ is the wave vector, and $\omega_{\rm p}=\sqrt{4\pi Q^2n/m}$ is the plasma frequency. Within the linear approximation, the screening length is expressed as a combination of ion and electron Debye radii,  $\lambda^{-2}=\lambda_i^{-2}+\lambda_e^{-2}$, where $\lambda_i=\sqrt{T_i/4\pi e^2 n_{i0}}$  and  $\lambda_e=\sqrt{T_e/4\pi e^2 n_{e0}}$, respectively. In laboratory gas discharges electrons are usually much hotter than ions ($T_e\gg T_i$) and screening is mostly associated with the ion component, $\lambda\simeq\lambda_i$. 

The last term in the right-hand-side of Eq.~(\ref{DAW}) corresponds to the pressure associated with the particle component:  $\mu=(1/T)(\partial P/\partial n)$ is the compressibility modulus and $v_{\rm T}=\sqrt{T/m}$ is the particle thermal velocity. Pressure and compressibility modulus include the corresponding contributions from the neutralizing medium. In the weakly coupled limit, when particle-particle correlations are absent, the pressure associated with interparticle interactions is canceled out exactly by the contribution from the neutralizing medium (ion-electron plasma)~\cite{KhrapakPRE2015_Sound,KhrapakPPCF2016}. The only remaining contribution to the pressure is the kinetic one, $P=nT$. Assuming a phenomenological equation of state of the form $P\propto n^{\gamma}$ we get $\mu=\gamma$, where $\gamma$ is the effective polytrope index (for instance, $\gamma=5/3$ for an adiabatic process in 3D and $\gamma=1$ for an isothermal process). The appropriate value for $\gamma$ in the ideal plasma regime will be identified below by means of the kinetic consideration. Expressed in reduced units the dispersion relation becomes
\begin{equation}\label{fluid}
\frac{\omega^2}{\omega_{\rm p}^2}= \frac{q^2}{q^2+\kappa^2}+\gamma\frac{q^2}{3\Gamma},
\end{equation}   
where $q=ka$. This expression is appropriate at sufficiently weak coupling. If $T\rightarrow 0$ ($\Gamma\rightarrow \infty$), the kinetic term vanishes (cold plasma limit). The resulting dispersion relation is usually referred to as the DAW dispersion relation~\cite{RaoDAW,FortovUFN,MerlinoPoP1998,MerlinoJPP2014}. At long wavelengths the acoustic dispersion is recovered, $\omega\simeq kc_{\rm{DA}}$ with $c_{\rm{DA}}=\omega_{\rm p}\lambda$. 


At sufficiently strong coupling the contribution from the neutralizing background dominates, and this makes $P$ amd $\mu$ negative~\cite{KhrapakPRE2015}. The sound velocity is therefore reduced compared to the weakly coupled value $c_{\rm DA}$~\cite{KalmanPRL2000,KhrapakPRE2015_Sound,KhrapakPPCF2016,KhrapakPoP2016,
SilvestriPRE2019,KhrapakPoP2019}. At strong coupling Eq.~(\ref{DAW}) becomes less and less accurate as $k$ increases, even when accurate values for the compressibility modulus are used~\cite{KhrapakISM}. Nevertheless, the acoustic asymptote $\omega = kc_{\rm s}$, with a properly evaluated sound velocity $c_{\rm s}$, can be used for sufficiently long wavelengths even at strong coupling~\cite{KhrapakPRE2015_Sound}.


\subsection{Kinetic description}

The dielectric permittivity of the multi-component isotropic collisionless Maxwellian plasma is~\cite{LL_Kinetics}
\begin{equation}
\epsilon(k,\omega)= 1 + \sum_{j}\frac{1}{k^2\lambda_j}\left[1+\frac{\omega}{\sqrt{2}kv_{{\rm T}j}}Z\left(\frac{\omega}{\sqrt{2}kv_{{\rm T}j}}\right)\right],
\end{equation}  
where the summation is over plasma components, $v_{{\rm T}j}=\sqrt{T_j/m_j}$ is the thermal velocity of the $j$-th component, $T_j$ and $m_j$ being the corresponding temperature and mass, and $Z(x)$ is the plasma dispersion function,
\begin{displaymath}
Z(x)=\frac{1}{\sqrt{\pi}}\int_{-\infty}^{+\infty}\frac{e^{-\xi^2}}{\xi-x}d\xi.
\end{displaymath}
Let us consider the three-component system, consisting of electrons, ions, and heavy highly charged particles (dust). For the low-frequency mode under consideration we have $\omega\ll kv_{{\rm T}_i}\ll kv_{{\rm T}e}$ and it suffices to retain only the static contribution from the ion and electron components. For the particle component and sufficiently long wavelengths $\omega \gg kv_{{\rm T}}$ the asymptotic ($x\gg 1$) expansion of $Z(x)$ reads~\cite{LL_Kinetics} 
\begin{displaymath}
Z(x)\simeq i\sqrt{\pi}e^{-x^2}-\frac{1}{x}\left(1+\frac{1}{2x^2}+\frac{3}{4x^4}+...\right).
\end{displaymath}  
Neglecting the exponentially small imaginary Landau damping term we arrive at
\begin{equation}
\epsilon(k,\omega)\simeq 1+\frac{1}{k^2\lambda^2}-\frac{\omega_{\rm p}^2}{\omega^2}\left(1+3\frac{k^2v_{\rm T}^2}{\omega^2}\right).
\end{equation} 
The dispersion relation is determined from the condition $\epsilon(k,\omega)=0$, which results in a quadratic equation for $\omega^2$. Its solution yields
\begin{equation}\label{kinetic}
\omega^2 \simeq \frac{\omega_{\rm p}^2 k^2\lambda^2}{1+k^2\lambda^2}+3k^2 v_{\rm T}^2.
\end{equation}
At short wavelengths $k\lambda\gg 1$, the (neglected) Landau damping term  will be large and the mode stops to propagate~\cite{LL_Kinetics}.

Comparing equations (\ref{kinetic}), (\ref{DAW}), and (\ref{fluid}) we conclude that they are equivalent if we set $\mu=\gamma=3$ in the fluid description. Thus, the kinetic description allows us to fix the numerical coefficient in the kinetic term. The term $3k^2 v_{\rm T}^2$ is often referred to as the Bohm-Gross term, after Ref.~\cite{Bohm}, and we follow this tradition here. Although the kinetic description is only applicable in the ideal plasma limit, when correlations between particles are absent (weak coupling), we will see below that the magnitude of the kinetic term itself is not changed much even at moderate (and possibly at strong) coupling.  

\subsection{Frequency moments and QLCA}  

The second frequency moments of the longitudinal and transverse current correlation functions are defined as~\cite{BalucaniBook}
\begin{equation}\label{Lmoment}
\omega_L^2(k)=3k^2 v_{\rm T}^2 +\frac{n}{m}\int\frac{\partial^2 \phi(r)}{\partial z^2}g(r)\left[1-\cos(\bf{kz})\right]d{\bf r}, 
\end{equation}
and 
\begin{equation}\label{Tmoment}
\omega_T^2(k)=k^2 v_{\rm T}^2 +\frac{n}{m}\int\frac{\partial^2 \phi(r)}{\partial x^2}g(r)\left[1-\cos(\bf{kz})\right]d{\bf r}. 
\end{equation}
The subscripts {\it L} and {\it T} refer to the longitudinal and transverse modes, respectively. The expressions above contain the kinetic (first term) and potential (or excess) contributions (second term). The potential contribution is expressed in terms of the pairwise interparticle interaction potential $\phi(r)$ and equilibrium radial distribution function (RDF) $g(r)$. The kinetic contribution to the longitudinal mode is formally given by the same Bohm-Gross term. 

The QLCA and QCA approximations~\cite{GoldenPoP2000,Hubbard1969,Takeno1971,KhrapakPoP2016} tell us that the dispersion relations of the longitudinal and transverse modes at strong coupling are given by the potential contributions in Eqs. (\ref{Lmoment}) and (\ref{Tmoment}). An exceptionally enlightening physical derivation demonstrating why it should be approximately so is due to Hubbard and Beeby~\cite{Hubbard1969}. Thus, QLCA approach does not take into account direct thermal effects. This is not a problem at strong coupling, because kinetic terms are numerically small in this regime. At weaker coupling kinetic effects should be accounted for and this should improve the accuracy of the QLCA~\cite{HouPRE2009,OttPRE2013}.   

The explicit expressions for $\omega_L(k)$ and $\omega_T(k)$ for the Yukawa interaction potential can be found elsewhere~\cite{KalmanPRL2000,DonkoJPCM2008,KhrapakPoP2016}. In the complete absence of correlations, for $g(r)=1$, the conventional DAW dispersion relation is recovered for the longitudinal mode
\begin{equation}
\omega_{L}^2=\frac{\omega_{\rm p}^2q^2}{q^2+\kappa^2}+3k^2v_{\rm T}^2,
\end{equation} 
which coincides with Eq.~(\ref{kinetic}). Note that the kinetic term vanishes at $T\rightarrow 0$ and we recover the conventional (cold plasma limit) DAW dispersion relation. This situation is, however, not internally consistent with the assumption of no correlations, $g(r)=1$.  

In the absence of correlations, the potential contribution to the transverse mode is identically zero, so that
\begin{equation}\label{T}
\omega_{T}^2 = k^2v_{\rm T}^2.
\end{equation}  
Even though it follows from Eq.~(\ref{T}) that the transverse frequency is non-zero due to the presence of the kinetic term, this mode is not supported in weakly coupled gases and moderately coupled fluids. It will not be considered further. Recent theoretical results regarding the onset and simple description of transverse waves in strongly coupled Yukawa fluids can be found elsewhere~\cite{KhrapakJCP2019,KhrapakIEEE2018}. The process of shear rigidity emergence with increasing coupling and inter-particle correlations in Yukawa systems, starting from the weakly coupled gaseous regime has been also investigated~\cite{KhrapakPoP2020}.  
 
\subsection{Generalized bulk modulus}

The dispersion relations resulting from the frequency moments and QLCA approaches can be also expressed in terms of generalized high-frequency (instantaneous) bulk ($K_{\infty}$) and shear ($G_{\infty}$) moduli as follows~\cite{NossalPR1968}:
\begin{equation}\label{Ginf}
\omega_T^2(k) = \frac{k^2}{mn}G_{\infty}(k), 
\end{equation}
and 
\begin{equation} \label{Kinf}
\omega_L^2(k) =\frac{k^2}{mn} \left[K_{\infty}(k) +\frac{4}{3}G_{\infty}(k)\right]. 
\end{equation}
In the long-wavelength limit ($k\rightarrow 0$),  $K_{\infty}$ and $G_{\infty}$ become just conventional instantaneous fluid elastic moduli~\cite{ZwanzigJCP1965,Schofield1966}.  The sum  $K_{\infty}+\tfrac{4}{3}G_{\infty}=M_{\infty}$ is known as the longitudinal modulus.  The essential physics behind the relevance of infinite frequency elastic moduli in the fluid regime is the following. If a perturbation is suddenly applied to a dense strongly coupled fluid (not too far from the fluid-solid phase transition), its initial response would not be very much different from that of a solid. It will respond elastically with the longitudinal response that depends on both the bulk and shear elastic moduli. This emphasizes the solid-like properties of strongly coupled fluids.

As the interparticle coupling weakens and the role of interparticle correlations diminishes, the transverse mode becomes irrelevant. It is tempting to assume that the longitudinal mode dispersion would decouple from the transverse one and depend on the generalized instantaneous bulk modulus alone, that is $\omega^2(k)\simeq \omega_L^2(k)-\tfrac{4}{3}\omega_T^2(k)$. In fact, similar conjecture has been demonstrated to result in a meaningful approximation for weakly and moderately coupled classical Coulomb fluids in two dimensions~\cite{KhrapakJCP2018}.
For a three-dimensional Coulomb fluid (one-component plasma), the dispersion relation of the form
\begin{equation}\label{bulkmod}
\omega^2=3k^2v_{\rm T}^2+\frac{k^2}{mn}\Delta K_{\infty}(k) 
\end{equation}  
has been demonstrated to capture correctly the onset of negative dispersion (the point where $d\omega/d k$ starts to be negative at $k\rightarrow 0$)~\cite{KhrapakPoP2016Onset}. Here $\Delta K_{\infty}$ is the excess component of the generalized bulk modulus. Motivated by the relative success of this approximation we have chosen to compare it with the results of MD simulations.      

\subsection{Strategy}

We adopt the following strategy. We have performed extensive MD simulations to determine the dispersion relation of the longitudinal collective mode for a broad parameter regime (see Fig.~\ref{Fig1}).  Direct comparison with the predictions of approximations described in this Section is used to test their relative success. In particular, we take three approximations: (i) weakly coupled expression from the fluid, kinetic, and frequency moments approaches, Eq.~(\ref{fluid}); (ii) second frequency moment of the longitudinal current correlation function (\ref{Lmoment}) keeping the Bohm-Gross kinetic term; (iii) expression (\ref{bulkmod}) based on the generalized instantaneous bulk modulus complemented with the Bohm-Gross kinetic term. The best choice among the considered approximations will be then identified.

\section{Numerical simulations}

The simulations were performed on graphics processing unit (NVIDIA Quadro P2000) using the HOOMD-blue software \cite{Anderson2008,Glaser2015}. We used $N=55 296$ Yukawa particles in a cubic box with
periodic boundary conditions. The cut-off radius for the potential has been chosen as $L_{\mathrm{cut}}=9\lambda_D$. The numerical time step was set to $\Delta t\simeq 2\times 10^{-3} \omega_p^{-1}$. The simulations were performed in the canonical $NVT$ ensemble with the Langevin thermostat at a temperature corresponding to the desired target coupling parameter $\Gamma$.

The system was first equilibrated for $1.25\times 10^6$ time steps with a drag coefficient $\gamma\simeq 2\times 10^{-2} \omega_p$. The drag coefficient was then reduced to $\gamma\simeq 2\times 10^{-4} \omega_p$ and the system was run for 
another 350 000 time steps. Finally the particle positions and trajectories were saved every 100 time steps for 180 000 time steps (except for $\kappa=4$ were the trajectories were saved every 400 time steps for 720 000 time steps in order to resolve 
more accurately the low frequency fluctuations of such system). 

The particle current was then calculated:
\begin{equation}
	\mathbf{J}({\bf k},t)=\sum_{j=1}^N \mathbf{v}_j(t)\exp(\imath \mathbf{k}\cdot \mathbf{r}_j(t)),
\end{equation}
where $\mathbf{v}_j (t)$ and $\mathbf{r}_j(t)$ are the velocity and position of the $j$-th particle and 
$\mathbf{k}$ is the wave vector. The Fourier transform in time was performed to obtain
the current fluctuation spectra. The particle positions were also used every 400 time steps to extract the accurate radial distribution functions $g(r)$.

To obtain the dispersion relation $\omega_l (k)$, the longitudinal current fluctuation spectrum $C_l (k,\omega)$ was fitted to the double-Lorentzian form \cite{KryuchkovSciRep2019,KhrapakJCP2018}:
\begin{equation}
	C_l(k,\omega)\propto \frac{\gamma_l(k)}{[\omega - \omega_l(k)]^2 + \gamma_l(k)^2}+\frac{\gamma_l(k)}{[\omega + \omega_l(k)]^2 + \gamma_l(k)^2},
\end{equation}
where $\gamma_l$ denotes the damping rate of the longitudinal mode.

\section{Results}  

We start by analysing the two very weakly coupled state points characterized by $\kappa=1$ and the two $\Gamma$ values, $\Gamma=0.2$ and $\Gamma=1.0$. For these state points the contribution from particle-particle interactions is very small and this gives us the opportunity to concentrate on the behaviour of the kinetic contribution to the dispersion relation.

\begin{figure}
\includegraphics[width=8.5cm]{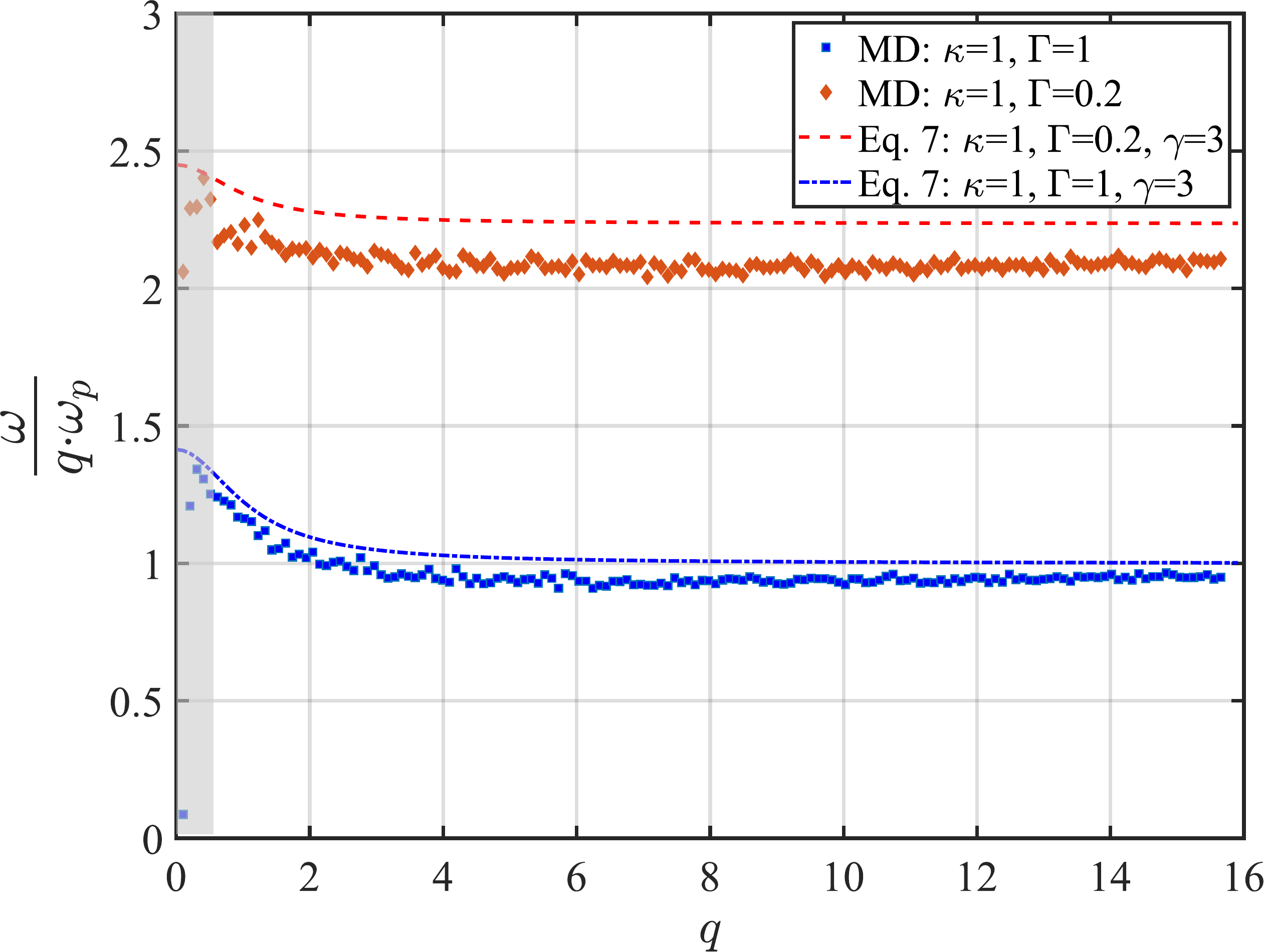}
\caption{(Color online) Ratio $\omega/q\omega_{\rm p}$ versus the reduced wave number $q=ka$ for the two weakly coupled state points, ($\kappa=1,\Gamma=0.2$) and ($\kappa=1,\Gamma=1$). The shaded area corresponds to the regime of insufficient statistics due to finite simulation volume. The curves correspond to Eq.~(\ref{fluid}) with $\gamma=3$.}
\label{Kinetic}
\end{figure}

The ratio $\omega/\omega_{\rm p}q$ versus $q$, in a very extended range of $q$, is plotted in Fig.~\ref{Kinetic}. From comparison with Eq.~(\ref{fluid}) we conclude that indeed the coefficient $\gamma$ is close to $3$, as the kinetic theory and frequency moments expression predict. Moreover, it remains constant in this very wide range of $q$. The region of very low $q$ is not well resolved, but here the reduced sound velocity seems to approach smoothly $\sqrt{6}\simeq 2.45$ for $\Gamma = 0.2$ and $\sqrt{2}\simeq 1.41$ for $\Gamma=1$, as Eq.~(\ref{fluid}) with $\gamma=3$ predicts. Thus, the Bohm-Gross terms $3k^2v_{\rm T}^2$ is appropriate in the range of $q$ investigated. The individual particle limit with $\omega^2\simeq 2k^2v_{\rm T}^2$~\cite{BalucaniBook,KryuchkovSciRep2019} is not reached in our simulations, see also Appendix.     

\begin{figure}
\includegraphics[width=7.5cm]{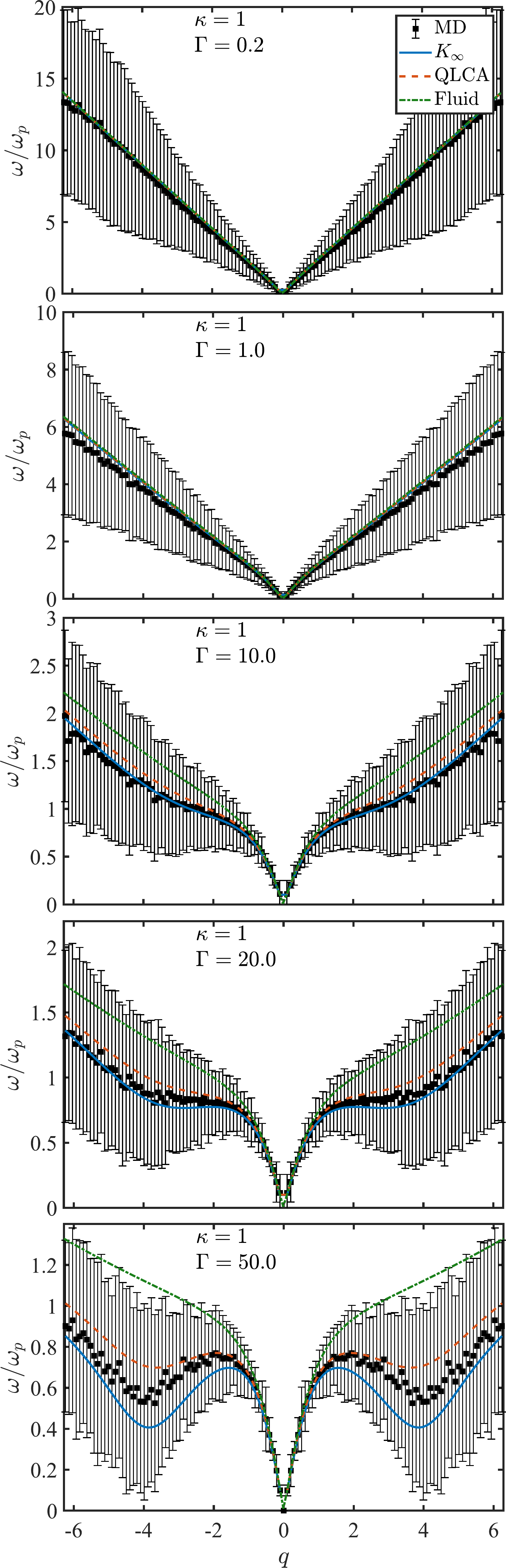}
\caption{(Color online) Dispersion relations of weakly coupled Yukawa fluids with $\kappa=1$. Symbols correspond to numerical results. Curves denote theoretical approximations compared in this work (see the legend).}
\label{FigK1}
\end{figure}

\begin{figure}
\includegraphics[width=7.5cm]{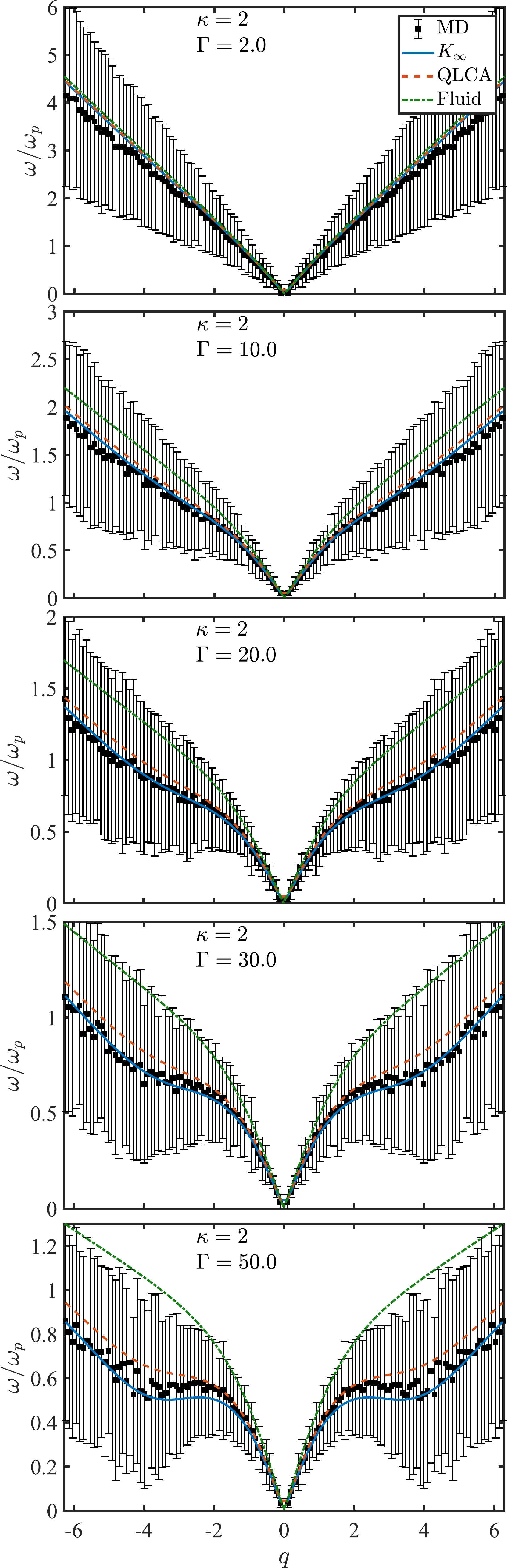}
\caption{(Color online) Dispersion relations of weakly coupled Yukawa fluids with $\kappa=2$. Notation is the same as in Fig.~\ref{FigK1}. }
\label{FigK2}
\end{figure}

\begin{figure}
\includegraphics[width=7.5cm]{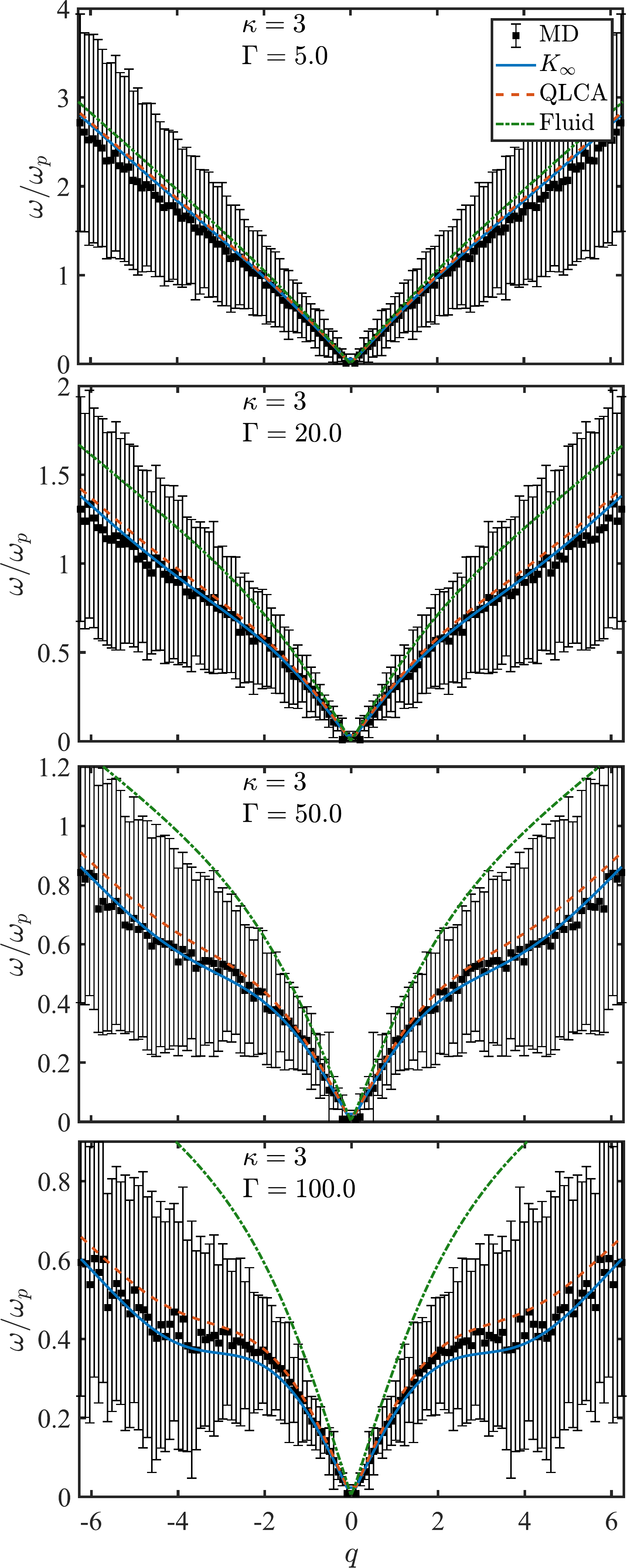}
\caption{(Color online) Dispersion relations of weakly coupled Yukawa fluids with $\kappa=3$.  Notation is the same as in Fig.~\ref{FigK1}.}
\label{FigK3}
\end{figure}

\begin{figure}
\includegraphics[width=7.5cm]{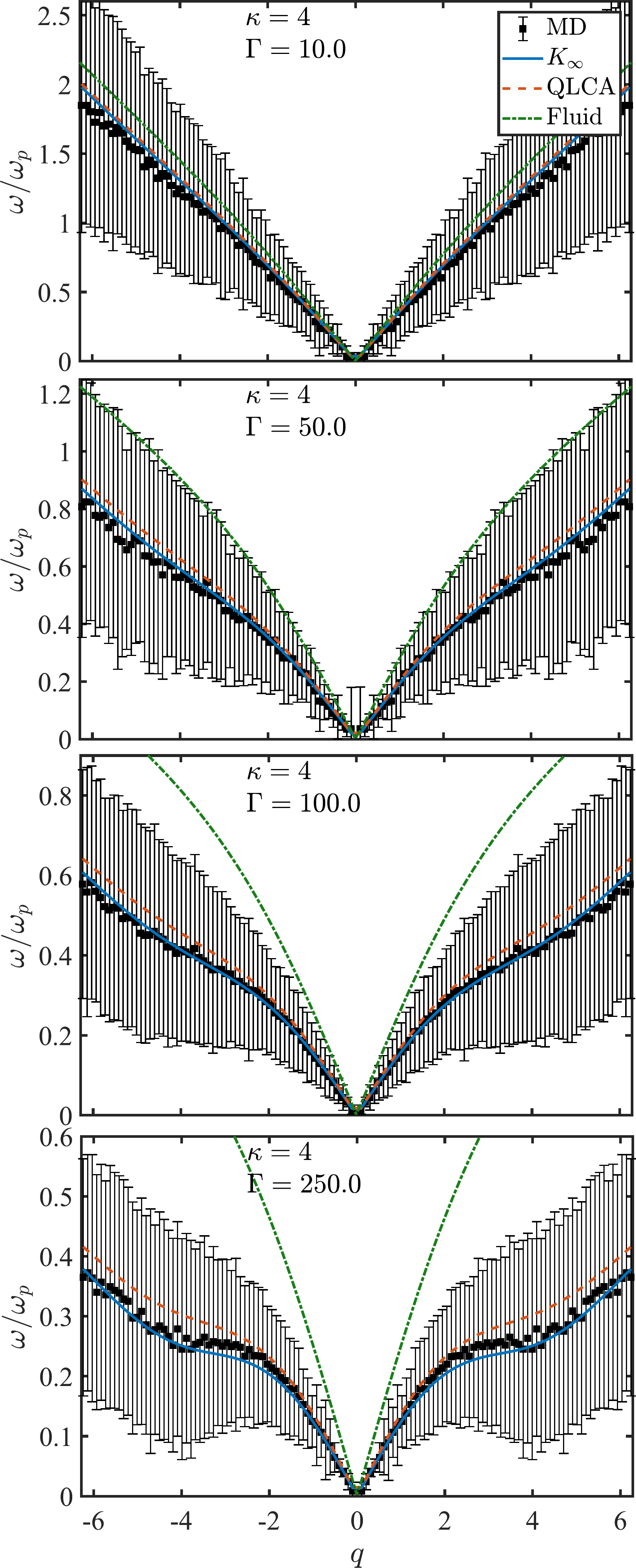}
\caption{(Color online) Dispersion relations of weakly coupled Yukawa fluids with $\kappa=4$.  Notation is the same as in Fig.~\ref{FigK1}.}
\label{FigK4}
\end{figure}

The dispersion relations obtained in our numerical experiment are shown in Figs.~\ref{FigK2} - \ref{FigK4} along with the theoretical curves used for the purpose of comparison. The following main trends can be summarized.  
   
In the weakly coupled regime all the theoretical approximations agree very well with the numerical data. Here the interparticle correlations are small and excess terms from the weakly coupled fluid and strongly coupled QCA approaches are nearly identical. There is also no difference between the generalized longitudinal and bulk moduli, because the excess shear modulus vanishes in the weakly coupled limit.


In the moderately coupled regime, the excess component of the generalized bulk modulus complemented by the Bohm-Gross kinetic term provides the best agreement with numerical data among the approximations considered. The second frequency moment expression (\ref{Lmoment}) somewhat overestimates the frequency.  

When approaching the strongly coupled regime, none of the approximations considered allow to describe numerical data accurately in the entire $q$-range investigated. The second frequency moment (\ref{Lmoment}) expression deviates to higher frequencies as $q$ increases. The generalized bulk modulus expression underestimate the frequency near the first minimum. This becomes particularly clear for the state point with $\kappa = 1$ and $\Gamma = 50$. Similar tendency has been observed in classical two-dimensional Coulomb fluids~\cite{KhrapakJCP2018}. In the long-wavelength regime, $q\lesssim 3$, the second frequency moment provides adequate description at strong coupling, as has been already noted in the Introduction. At even longer wavelengths ($q\ll 1$) the difference between the second frequency moment and generalized bulk modulus expressions practically disappears, because of the inequality $G_{\infty}\ll M_{\infty}\simeq K_{\infty}$ and acoustic character of the dispersion.     

The appearance of the minimum in the dispersion relation can serve as a pragmatic demarcation between the moderately and strongly coupled regimes.
This can have relations with the crossover between gas-like and fluid-like behaviour, the concept known as the ``Frenkel line'' on the phase diagram~\cite{BrazhkinPRE2012}.  

Regarding the weakly coupled fluid and kinetic approximations, they become particularly inappropriate when the screening parameter increases. This could be expected. Strong coupling effects are known to affect only weakly the magnitude of the sound velocity at $\kappa\lesssim 1$, but lead to its considerable decrease for higher $\kappa$~\cite{KalmanPRL2000}, see in particular Fig.~4 from Ref.~\cite{KhrapakPRE2015_Sound} and Fig.~5 from Ref.~\cite{KhrapakPPCF2016}.

\section{Conclusion}     

In this paper we have addressed the question regarding what determines the dispersion relations of Yukawa fluids at moderate coupling, when kinetic and potential contributions to the dispersion relations are of similar magnitude. Three theoretical approaches have been compared with the results from extensive MD simulations. Among these, an empirical expression combining the generalized excess bulk modulus with the Bohm-Gross kinetic term provides the best agreement with numerical results at weak and moderate coupling. The approach to the strong coupling regime is signalled by the appearance of pronounced minimum in the dispersion relation. In this regime none of the approximations considered allows to describe numerical data accurately in the entire range of wave vectors.

\acknowledgments
We thank Hubertus Thomas for reading the manuscript.
L. Cou\"edel acknowledges the support of the Natural Sciences and Engineering Research Council of Canada (NSERC), RGPIN-2019-04333.

\appendix*
\section{Free-particle limit}

\begin{figure}
\includegraphics[width=7.5cm]{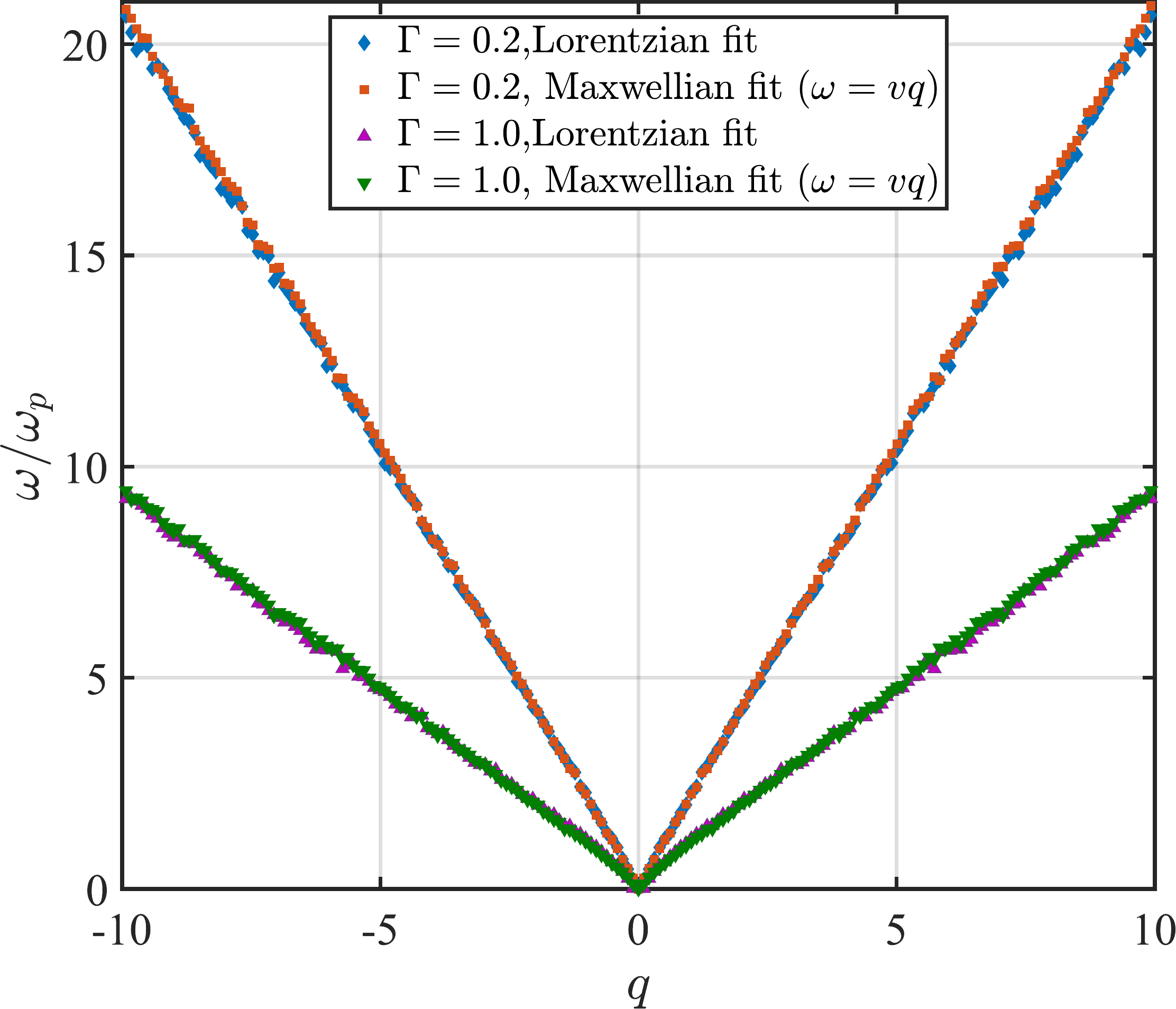}
\caption{(Color online) Reduced frequency $\omega/\omega_{\rm p}$ versus the reduced wave number $q=ka$ for the two weakly coupled state points, ($\kappa=1,\Gamma=0.2$) and ($\kappa=1,\Gamma=1$). The results of analysis using double-Lorentzian and Maxwellian fits are almost indistinguishable.}
\label{Fig5}
\end{figure}

In the limit of very large wavevectors, the current fluctuations spectra are expected to approach their free-particle limiting expressions. For the longitudinal spectrum this corresponds to the Maxwellian shape~\cite{BalucaniBook,KryuchkovSciRep2019}
\begin{displaymath}
C_l(k,\omega)\propto \left(\frac{\omega}{q}\right)^2\exp\left(-\frac{m\omega^2a^2}{2Tq^2}\right),
\end{displaymath}  
which is peaked at $\omega a=\pm \alpha q v_{\rm T}$ with $\alpha=\sqrt{2}$. We have repeated the analysis of the two weakly coupled state points longitudinal current spectra using the Maxwellian shape and treating $\alpha$ as a free parameter. No significant difference from the double-Lorentzian form is evident, see Fig~(\ref{Fig5}). The coefficient $\alpha$ appears closer to $\sqrt{3}$ than to $\sqrt{2}$. The free-particle limit seems not reached in our simulation.   

\bibliographystyle{aipnum4-1}
\bibliography{YukawaGas}

\end{document}